\documentclass[conference]{IEEEtran}
\IEEEoverridecommandlockouts
\usepackage{cite}
\usepackage{amsmath,amssymb,amsfonts}
\usepackage{bbm}
\usepackage{algorithmic}
\usepackage{graphicx}
\usepackage{textcomp}
\usepackage{xcolor}
\usepackage{dsfont}
\usepackage{physics}
\usepackage{comment}
\usepackage{hyperref}
\usepackage[acronym]{glossaries}
\def\BibTeX{{\rm B\kern-.05em{\sc i\kern-.025em b}\kern-.08em
    T\kern-.1667em\lower.7ex\hbox{E}\kern-.125emX}}

\counterwithin{equation}{section}
\newcounter{thm}
\newtheorem{theorem}[thm]{Theorem}

\newtheorem{definition}[thm]{Definition}

\newtheorem{lemma}[thm]{Lemma}

\newcommand{\eins}{\mathbbm{1}}

\newacronym{povm}{POVM}{Positive Operator Valued Measurement}
\newacronym{csi}{CSI}{Channel State Information}

\begin{document}

\title{Capacity Formulas for the Lossy Bosonic Compound Wiretap Channel\\

\thanks{This work was financed by the DFG via grant NO 1129/2-1 and by the BMBF via grants 16KISQ039 and 16KISQ077. The authors acknowledge the financial support by the Federal Ministry of Education and Research of Germany in the programme of “Souverän. Digital. Vernetzt.”. Joint project 6G-life, project identification number: 16KISK002. Further, this work was supported by the state of Bavaria via the 6GQT project.}}

\author{\IEEEauthorblockN{Florian Seitz, Janis Nötzel, \emph{Member, IEEE}}
\textit{Emmy-Noether Group Theoretical Quantum Systems Design},\\
\textit{Technical University of Munich, Munich, Germany},\\
\textit{\{flo.seitz,janis.noetzel\}@tum.de}}

\maketitle
\begin{abstract}
    We consider the bosonic compound wiretap channel. A pair of lossy channels connects a sender with both a (legitimate) receiver and an eavesdropper. The sender and receiver have only partial information about the actual state of the channels. In this situation, their task is to transmit the maximum amount of messages over an asymptotically large number of uses of their channel while guaranteeing at the same time that only an (asymptotically) negligible amount of information leaks to the eavesdropper. We prove capacity formulas for the case where both sender and receiver have the same information about the system and for the case where the sender has channel state information.
\end{abstract}

\section{Introduction}
 Communication over a compound wiretap channel demands correct transmission to the receiver, independently of the channel conditions, which are allowed to take on arbitrary values within some pre-defined set. In practical scenarios, such situations can be of importance when the transmitter has little information about the channel state and low latency data transmission must be ensured so that a strategy of sending pilot symbols that allows the receiver to learn the channel and then transmit its information back to the sender is ruled out. Due to the digitization of more and ever more services, data security is an increasingly relevant topic, which is reflected in physical layer communication system research via the presence of an eavesdropper. It must be assumed that a lack of information about the channel conditions connecting the sender to the eavesdropper is present as well, which increases the challenge of guaranteeing not only accurate data transmission to the legitimate receiver, but also vanishing information leakage to the eavesdropper.   

 The corresponding classical system has been investigated in works such as \cite{wyner} when all channel parameters are perfectly known, and in \cite{kramerCompound,bbs} for compound wiretap channels. 
 The secrecy capacity of quantum channels with finite-dimensional Hilbert spaces was proven in \cite{devetak} and independently in \cite{Cai2004QuantumPA}. The private capacity of bosonic systems was further investigated in \cite{jeongPrivateBosonic,PirandolaPrivateBosonic,sharmaWildePrivateBosonic}.

\section{Notation and System Model}
The Fock space is denoted $\mathcal H$, and the set of all density operators on it is $\mathcal{P}(\mathcal{H})$. The von-Neumann Entropy is written as $S$, the entropy of a thermal state with an average of $E$ thermal photons is $g(E)$ where $g(x) = (x+1) \log(x+1) - x \log(x)$ is the Gordon function \cite{holevoBOOK}. The binary entropy is $h:[0,1] \to [0,1]; x \mapsto -x\log x - (1-x) \log(1-x)$. Throughout, the logarithm has base two.
The Holevo quantity of an ensemble $\{p_i,\rho_i\}_i$ is $\chi(\{p_i,\rho_i\}_i)$.
In this work, we focus on noiseless bosonic classical-quantum channels, which are defined by one parameter $\tau\in[0,1]$, where $\tau^2$ is called the transmissivity and $1-\tau^2$ is the loss. Upon receiving the input $\alpha\in\mathbb C$, the channel output is given by $\mathcal N_\tau(\alpha)=|\tau\alpha\rangle\langle\tau\alpha|$. We use the convention that Greek letters indicate coherent states, so that $|\tau\alpha\rangle=e^{-|\tau\alpha|^2/2}\sum_n\tfrac{(\tau\alpha)^n}{\sqrt{n!}}|n\rangle$ is a coherent state, while $|n\rangle$ are so-called \emph{photon number states} which form an orthonormal basis of $\mathcal H$. A noiseless bosonic compound wiretap channel is then given by a set $\mathcal N=\{\mathcal N_{\tau},\mathcal 
N_{\eta}\}_{(\tau,\eta)\in\mathbf S}$ of channels where $\mathbf S$ is a subset of $[0,1]\times[0,1]$, $\tau$ quantifies the transmissivity of the transmission link between the legal parties and $\eta$ that of the link from the legal sender to the wiretapper.
Throughout, $\mathcal X$ is a set, and $X$ denotes a random variable on it. $\mathbb{E}_X$ denotes the expectation value with respect to $X$. For a real number $r$, $(r)_+=r$ if $r>0$ and $(r)_+=0$ else.
\begin{definition}[$(n,M_n,\lambda,\mu)$ Code] \label{def:code_csi}
    A collection $\{x_{s,m}\}_{m\in[M_n],s\in \mathbf S}$ of signals where each $x_{s,m}$ is an element of $\mathbb C^n$ and a \gls{povm} $\{D_m\}_{m=1}^{M_n}$ are called an $(n,M_n,\lambda,\mu)$ code $\mathcal C$ with state information at the transmitter if for all pairs $s = (\tau, \eta) \in \mathbf S$ the success probability 
    \begin{align}\label{eq:p_suc_def}
        p_{\mathrm{suc},s}(\mathcal C):=\frac{1}{M_n}\sum_{m=1}^{M_n}\Tr[D_m \mathcal N_\tau^{\otimes n}(x_{s,m})]
    \end{align}
    satisfies $p_{\mathrm{suc},s}(\mathcal C)\geq1-\lambda$ and the information leakage to the eavesdropper satisfies
    \begin{align} \label{eq:privacy}
        \chi(\{M_n^{-1},\mathcal N_\eta^{\otimes n}(x_{s,m})\}_{x_m=1}^{M_n}) < \mu. 
    \end{align}
    If the signals $\{x_{m}\}_{m\in[M_n]}$ do not depend on $s$ and for all $s = (\tau, \eta) \in \mathbf S$ it holds $p_{\mathrm{suc},s}(\mathcal C)\geq1-\lambda$ and
    \begin{align}
        \chi(\{M_n^{-1},\mathcal N_\eta^{\otimes n}(x_{m})\}_{x_m=1}^{M_n}) < \mu 
    \end{align}
    the code is called an $(n,M_n,\lambda,\mu)$ code $\mathcal C$ without state information at the transmitter.
\end{definition}

\begin{definition}[Achievable Rates, Capacity]
    A rate $R\geq0$ is called achievable with/without state information at the transmitter for the classical-quantum compound wiretap channel $\mathcal N$ under state constraint $\mathcal R$ if there exists a sequence $(\mathcal C_n)_{n\in\mathbb N}$ of $(n,M_n,\lambda_n,\mu_n)$ codes with/without state information at the transmitter, obeying the state constraint $\mathcal R$, such that both $\lambda_n\to0$ and $\mu_n\to0$ and $\liminf_{n\to\infty}\frac{1}{n}\log M_n\geq R$. 
    
    The secret message transmission capacity of $\mathcal N$ with (without) state information at the transmitter is then defined as the supremum over all rates that are achievable for $\mathcal N$ with (without) state information and is denoted as $C_\text{CSI}$ ($C_\text{noCSI} $).
\end{definition}
\section{Preliminaries}

\begin{lemma}[Covering Lemma {\cite[Chapter 17]{wilde2017}}] \label{thm:covering_lemma}
Let $\mathcal X$ be a set and $p_{X}(x)$ a probability distribution over $\mathcal X$. Then $\left\{ p_{X}(x), \rho_{x} \right\}_{x \in \mathcal X}$, where $\left\{\rho_{x}\right\}_{x \in \mathcal X} \subset \mathcal P(\mathcal H)$ for some $\mathcal{H}$, is called the true ensemble. Let $\mathcal L$ be a set and select a random code word $X_l$ according to $p_{X}(x)$ for each $l \in \mathcal L$, then $\left\{ \frac{1}{\left| \mathcal L \right|},\rho_{X_l} \right\}_{l \in \mathcal L}$ is called the fake ensemble. Let there be a code space projector $\Pi$ that is an orthogonal projector satisfying
\begin{align}
\mathrm{Tr}\left[\rho_{x} \Pi \right] \geq 1-\varepsilon, \qquad \mathrm{Tr}\left[\Pi\right] \leq D,
\end{align}
and a set of codeword projectors $\left\{ \Pi_{x} \right\}_{x \in \mathcal X}$ with the properties
\begin{align}
\mathrm{Tr}\left[\rho_{x}\Pi_{x}\right] \geq 1-\varepsilon, \qquad
\Pi_{x} \rho_{x}\Pi_{x} \leq \frac{1}{d}\Pi_{x},
\end{align}where $\varepsilon > 0$ and $0<d<D$. Let $\bar{\rho} = \sum_{x \in \mathcal X}p_{X}(x) \rho_{x}$ be the average state of the true ensemble and let $\bar{\rho}_{\mathcal L} = \frac{1}{|\mathcal L|}\sum_{l \in \mathcal L} \rho_{s}$ be the average state of the fake ensemble. Then
\begin{align}
\mathrm{Pr}\left\{ \|\bar{\rho}-\bar{\rho}_{\mathcal L}||_{1} \leq 30\varepsilon^{\frac{1}{4}} \right\} &\geq 1-2D e^{-\frac{\varepsilon^{3}|\mathcal L|d}{4 D}},
\end{align}
given $\varepsilon$ is small and $|\mathcal L| \gg \frac{\varepsilon^{3}d}{D}$. The probability is with respect to the random codebook selection. Note that the original bound is a bit tighter.
\end{lemma}
We also make extensive use of the concepts of quantum typicality and strong classical typicality. Key properties are listed below; a full discussion is available in \cite{wilde2017}.
\begin{definition}[Strong Typicality]
Let $\mathcal X$ be a finite set, $x^n = (x_1, \dots, x_n)$ a sequence, $x_1,\ldots,x_n\in\mathcal X$ and $N(x|x^n)$ the number of times the symbol $x$ occurs in $x^n$. Let $p_X(x)$ be a probability distribution for a random variable $X$. For $\delta > 0$, the $\delta$-strongly typical set $T_\delta^{X^n}$ consists of all $x^n$ where
\begin{align}
    \forall x \in \mathcal{X}, 
\begin{cases}
\left| \frac{1}{n} N(x|x^n) - p_X(x) \right| \leq \delta & \text{if } p_X(x) > 0,\\
\frac{1}{n} N(x|x^n) = 0 & \text{else}
\end{cases}.\nonumber
\end{align}
For $\varepsilon, \delta > 0$, sufficiently large $n$ and a positive constant $c$, $T_{\delta}^{X^n}$ has the properties
\begin{align}
    \text{Pr}\{X^n \in T_\delta^{X^n}\} &\geq 1-(2n)^{-|\mathbf X|}2^{-n\varepsilon^2\log(2)/2}\\
    (2n)^{-|\mathbf X|} 2^{n (H(X) - c \delta)} &\leq |T_\delta^{X^n}|\leq 2^{n (H(X) + c \delta)},
\end{align}
where $H(X) = \sum_{x \in \mathcal X}p_X(x)\log(p_X(x))$ is the entropy.
\end{definition}
\begin{definition}[Typical Subspace]
Let $\rho \in \mathcal{P}(\mathcal{H})$ be a density operator with the spectral decomposition $\rho = \sum_{x\in\mathcal X} p_X(x) \ketbra{x}$. For every $n \in \mathbb{N}$, the projector onto the typical subspace of $\rho^{\otimes n}$ is defined as
\begin{align}
\Pi_{\rho, \delta}^n = \sum_{x^n \in T_\delta^{X^n}} \ketbra{x^n},
\end{align}
where $T_\delta^{X^n}$ is the set of typical sequences of length $n$ for $p_{X^n}(x^n) = p_X(x_1) \dots p_X(x_n)$.
It has the following properties for sufficiently large $n$ and any $\epsilon, \delta > 0$:
\begin{align}
\mathrm{Tr}\left[\Pi_{\rho, \delta}^n\right] &\leq 2^{n \left( S(\rho) + \delta \right)}, \\
\mathrm{Tr}\left[\Pi_{\rho, \delta}^n \rho^{\otimes n}\right] &\geq 1 - \epsilon, \\
2^{-n\delta} \Pi_{\rho, \delta}^n &\leq 2^{n S(\rho)}\Pi_{\rho, \delta}^n \rho^{\otimes n} \Pi_{\rho, \delta}^n \leq 2^{n\delta} \Pi_{\rho, \delta}^n.
\end{align}
The space $\mathrm{span}\left\{ \ket{x^n} \right\}_{x^n \in T_\delta^{X^n}}$ is called the typical subspace.
\end{definition}
\begin{lemma}[Coherent State Approximation]\label{thm:coher_approx}
Let $\ket{\alpha}$ be a coherent state and let $P_{N} = \sum_{n=0}^N \ketbra{n}$, where $\ket{n}$ are photon number states, be the projector onto a Fock space that is truncated at $N$. Then
\begin{align}
\mathrm{Tr}\left[P_{N} \ketbra{\alpha} \right] \geq 1-\frac{1}{2} 2^{-N}
\end{align}
for all $N > 8e|\alpha|^2$. Clearly the same bound applies to a state $\rho=\sum_{\alpha \in A} p(\alpha) \ketbra{\alpha}$ as long as $N > 8e|\alpha|^2 \; \forall \alpha \in A$.
\end{lemma}
Lemma \ref{thm:coher_approx} follows from Stirling's formula and the observation that for every $c>0$ we have $(c/N)^N<2^{-N}$ once $N$ is large enough.
\begin{lemma} \label{thm:compound_rate}
Let $\mathcal{X}$ be a finite set, $\mathcal{H}$ a Hilbert space, and $\mathcal{N} = \{\mathcal{N}_s\}_{s \in \mathbf{S}} \subset C(\mathcal{X}, \mathcal{H})$ a classical-quantum compound channel, where $\mathbf{S}$ is a finite set of channel states. Let $\{\ket{e_x}\}_{x \in \mathcal{X}}$ be an orthonormal basis of the complex Hilbert space $\mathbb{C}^{|\mathcal{X}|}$, and let $p(x)$ for $x \in \mathcal{X}$ be a probability distribution over $\mathcal{X}$. Furthermore we define the "pruned" distribution $p'_{X'^n}$ as
\begin{align}\label{eq:pruned_dist}
    p'_{X'^n}(x^n) = \frac{1}{\Delta_{\delta}^n} p_{X^n}(x^n)\eins_{T_{\delta}^{X^n}}(x^n)
\end{align}
where $\Delta_{\delta}^n = \sum_{x^n \in T_{\delta}^{X^n}} p_{X^n}(x^n)$.
For block length $n \in \mathbb{N}$, define the following states on $\mathbb{C}^{|\mathcal{X}|} \otimes \mathcal{H}$:
\begin{align}
\rho_{n} =& \frac{1}{\left| \mathbf{S} \right| }\sum_{s \in \mathbf{S}} \sum_{x^n \in \mathbf{X}^n} p_{n}(x^n) \ketbra{e_{x^n}} \otimes \mathcal N_{s}^{\otimes n}(x^n), \label{eq:rho_n}\\
\tau_{n} =& \left( \sum_{x^n \in \mathbf{X}^n} p_{n}(x^n) \ketbra{e_{x^n}}   \right) \nonumber \\ & \otimes \frac{1}{\left| \mathbf{S} \right| }\sum_{s \in \mathbf{S}} \left( \sum_{x^n \in \mathbf{X}^n} p_{n}(x^n)\mathcal N_{s}^{\otimes n}(x^n) \right).\label{eq:tau_n}
\end{align}
Let $\mathcal C^{n} = \left\{C_m^{n}\right\}_{m =1}^{K_{n}}$ be a random code, where every element $C_m^{n}$ is selected randomly and independently from $\mathcal X^n$ according to $p'_n(x)$. Then if for some $\lambda \in [0, 1]$ and $a > 0$ there exists a projector $P_n$ such that
\begin{align}
\mathrm{Tr}[P_n \rho_n] \geq 1 - \lambda \qquad \text{and} \qquad \mathrm{Tr}[P_n \tau_n] \leq 2^{-n a},
\end{align}
for any $\gamma$ with $0 < \gamma \leq a$ and $K_n = \lfloor 2^{n(a-\gamma)} \rfloor$, then there exists a decoding \gls{povm} $\{D_{C_m^n}^n\}_{m = 1}^{K_n}$ 
such that
\begin{align}
\mathbb P&(\min_{s \in \mathbf{S}} p_{err}(\mathcal C)<\xi^{\frac{1}{2}})\geq1-\xi^{\frac{1}{2}}
\end{align}
where 
\begin{align}
    p_{err}=&1-\frac{1}{K_n} \sum_{m=1}^{K_n} \mathrm{Tr} \big[\mathcal{N}_s^{\otimes n}(C_m^n)D_{C_m}^n\big], \text{ and}\\
    \xi =& |\mathbf{S}|  (2 \lambda + 2(1-\Delta^n_{\delta}) + \frac{4}{(\Delta_{\delta}^{n})^{2}} 2^{-n \gamma })
\end{align}
\end{lemma}
This lemma is an adapted version of Lemma 1 from \cite{Bjelakovic2013}, which also contains
\begin{lemma}[{\cite[Lemma 5]{Bjelakovic2013}}\label{thm:compound_projector}]
For every $\delta > 0$, finite compound cq-channel $\mathcal{N} = \left\{ \mathcal{N}_{s} \right\}_{s \in \mathbf{S}} \subset C(\mathcal{X}, \mathcal{H})$ and probability distribution $p$ over $\mathcal{X}$, there exists a constant $\tilde{c}$ such that for every sufficiently large $n \in \mathbb{N}$, there exists a projector $P_{n,\delta} \in \mathcal{B}\left((\mathbb{C}^{\left| \mathcal{X} \right|})^{\otimes l} \otimes \mathcal{H}^{\otimes l}\right)$ with the properties
\begin{align}
    \mathrm{Tr}\left[P_{n,\delta} \rho_{n}\right] &\geq 1 - \left| \mathbf{S} \right| \, 2^{-n\tilde{c}}, \label{eq:code_const_proj_prop1} \\
    \mathrm{Tr}\left[P_{n,\delta}\tau_n\right] &\leq 2^{-n(a-\delta)}, \label{eq:code_const_proj_prop2}
\end{align}
where $a := \min_{s \in \mathbf{S}} D(\rho_{s} \| \hat{p} \otimes \sigma_{s})$ with the states
\begin{align}
    \hat{p} :=& \sum_{x \in \mathbf{X}}p(x)\ketbra{e_{x}},\ \ \ \sigma_{s} := \sum_{x \in \mathbf{X}}p(x) \mathcal N_{s}(x),\\
    \rho_{s} :=& \sum_{x \in \mathbf{X}}p(x)\ketbra{e_{x}} \otimes \mathcal N_{s}(x).
\end{align}
$D(\,\cdot\, \|\,\cdot\,)$ is the Quantum Relative Entropy and $\rho_n$, $\tau_n$ are defined according to \eqref{eq:rho_n}, \eqref{eq:tau_n}.
\end{lemma} 
We note that the proof of Lemma \ref{thm:compound_projector} is valid also in cases where $|\mathbf S|=\mathcal O(n^t)$ for some $t>0$, as can be seen explicitly in \cite[Eq. (74)]{Bjelakovic2013}.
\begin{lemma}[Finite Support Approximation] \label{thm:fin_sup}\\
    Let $0 \leq \rho, \sigma, \Lambda \leq \mathds{1}$. Then
    \begin{align}
    \Tr[\Lambda \rho] \leq \Tr[\Lambda \sigma] + \| \rho - \sigma \|_1.
    \end{align}
A proof is given in \cite{wilde2012}.
\end{lemma}
\begin{lemma}[Von-Neumann Entropy Continuity Bound]\label{thm:entropy_continuity}
Let $\hat{N}$ denote the photon number operator, and consider two quantum states $\rho$ and $\sigma$ on an infinite-dimensional separable Hilbert space $\mathcal{H}$. Suppose these states satisfy the conditions
\begin{align}
\max\{\mathrm{Tr}\left[\hat{N} \rho\right],\mathrm{Tr}\left[\hat{N} \sigma\right]\} \leq E \leq \infty,\ \mathrm{and}\  
\frac{1}{2} \left\| \rho - \sigma \right\|_1 \leq \varepsilon,\nonumber
\end{align}
where $0 \leq \varepsilon \leq \frac{E}{1 + E}$. Under these conditions, the von Neumann entropies of $\rho$ and $\sigma$ satisfy the continuity bound
\begin{align}
\left| S(\rho) - S(\sigma) \right| \leq h(\varepsilon) + E \cdot h\left( \frac{\varepsilon}{E} \right).
\end{align}
The version is from \cite{Becker2021}, based on \cite{Winter2015}.
\end{lemma}

\section{Results}
\begin{theorem}\label{thm:main}
    For the compound wiretap channel, defined by the state set $\mathbf{S}$ satisfying $(\tau,\eta)\in\mathbf S\implies\tau>\eta$ and with an average power constraint requiring that for every block length $n$ and each code of length $M_n$ the respective code words fulfill $\sum_{i = 1}^{M_n} \abs{x_i}^2 \leq nE$, the secret message transmission capacity with \gls{csi} at the sender is
    \begin{align}
        C_\text{CSI} = \inf_{s \in \mathbf{S}} \left( g(\tau E) - g(\eta E) \right),
    \end{align}
    and without \gls{csi} it is
    \begin{align}
        C_\text{noCSI} = \big(\inf_{s \in \mathbf{S}} g(\tau E) - \sup_{s \in \mathbf{S}} g(\eta E)\big)_+ .
    \end{align}
\end{theorem}
\section{Proof}
\emph{Converse:} Both converse results follow from the converse for the bosonic wiretap channel \cite[Theorem 27]{sharmaWildePrivateBosonic}. In the case of state information, the worst-case choice over all $(\tau,\eta)\in\mathbf S$ dictates the performance. If no such state information is provided, the two requirements for (a) correct and (b) secure data transmission decouple and \cite[Theorem 27]{sharmaWildePrivateBosonic} again provides the converse.

\emph{Direct Part:} We begin by considering a finite set of channel states $\mathbf{S}_n$ which we let depend on the block-length $n$ as $|\mathbf S_n|=\mathcal O(n^t)$ for some $t\geq0$ and a finite $\mathcal{X} \subset \mathbb{C}$ and later extend the result to continuous  sets. Let $(\tau,\eta) = s \in \mathbf{S}_n$, where $\tau$ is the transmissivity of the channel to the receiver and $\eta$ the transmissivity to the wiretapper. Consider sequences $x^n = x_{1}, \dots, x_{n}$, where each $x_{i}$ takes values from $\mathcal{X}$. Then with $n$ uses of the channel the receiver gets the state $\rho^{s}_{x^n} = \rho^{s}_{x_{1}}\otimes\dots \otimes \rho^{s}_{x_{n}}$ and the wiretapper gets $\tilde{\rho}^{s}_{x^n}=\tilde{\rho}^{s}_{x_{1}}\otimes\dots \otimes\tilde{\rho}^{s}_{x_{n}}$, where $\rho^{s}_{x}=\ketbra{\tau x}{\tau x}$ and $\tilde{\rho}^{s}_{x}=\ketbra{\eta x}{\eta x}$.

Let $X^n$ be the a random variable with the $n$-fold probability distribution $p_{X^n}(x^n) = p_{X}(x_{1})  \dots  p_{X}(x_{n})$ and let $p'_{X'^n}$ be the corresponding ''pruned`` distribution, as defined in \ref{eq:pruned_dist}. We consider finite dimensional approximations of the transmitted states using lemmas \ref{thm:fin_sup} and \ref{thm:coher_approx}. For a state $\rho$ we set $\rho' = \lambda\cdot P_{N} \rho P_{N}$, 
where $\lambda>0$ ensures normalization. Based on this definition and lemma \ref{thm:coher_approx} we get $\|\rho-\rho'\|_{1}\leq 2^{-N}$ for sufficiently large $N$. For $n$-mode states, we have to multiply this bound by $n$, which implies that to get a vanishing error for large $n$, we need $N \sim \log(n)$, meaning that our $n$ mode approximated state has $\sim \log(n)^n$ dimensions. If we set for example $N = 2 \log(n)$, the maximum approximation error of an $n$ mode state is upper bounded by $\frac{1}{n}$ for sufficiently large $n$. For both the capacity and the secrecy parts, we first show that good codes exist for the approximated states, and then argue that the same codes and operators perform almost as well with the original states.

Let $\Bar{\rho}'^s = \mathds E_{X} \rho'^s_{X} = \sum_{x \in \mathcal{X}} p_{X}(x) \rho'^{s}_{x}$ be the finitely approximated single mode average state at the receiver. Set $a = \inf_{s \in \mathbf S_n} S(\Bar{\rho}'^s)$, take a $\gamma$ with $0 < \gamma \leq a$ and define $K_n = \lfloor 2^{n(a-\gamma)} \rfloor$. Then lemmas \ref{thm:compound_rate} and \ref{thm:compound_projector} show that for sufficiently large $n$, a random code $\mathcal C^{n} = \left\{C_m^{n}\right\}_{m =1}^{K_{n}}$ exists where every element $C_m^{n}$ is selected randomly and independently from $\mathcal X^n$ according to $p'_n(x)$, with corresponding \gls{povm} $\{D_{C_m^{n}}^n\}_{m = 1}^{K_n}$ and associated constant $\tilde c$ such that
\begin{align}
&\mathbb{E}_{\mathcal{C}^n } \min_{s \in \mathbf{S}_n} \frac{1}{K_n} \sum_{m=1}^{K_n} \mathrm{Tr} \big[\rho'^s_{C_m^n} D_{C_m^{n}}^n\big] \nonumber\\ \geq& 1 - |\mathbf{S}_n|  (2 |\mathbf{S}_n| 2^{-n\tilde{c}} + 2(1-\Delta^n_{\delta}) + \frac{4}{(\Delta_{\delta}^{n})^{2}} 2^{-n \gamma }),
\end{align}
and thus with lemma \ref{thm:fin_sup} for $n$ large enough so that $(\Delta_{\delta}^{n})^{2}>1/2$
\begin{align}
&\mathbb{E}_{\mathcal{C}^n } \min_{s \in \mathbf{S}_n} \frac{1}{K_n} \sum_{m=1}^{K_n} \mathrm{Tr} \big[\rho^s_{C_m^n} D_{C_m^{n}}^n\big] \nonumber\\ \geq& 1 - |\mathbf{S}_n|  (2 |\mathbf{S}_n| 2^{-n\tilde{c}} + 2(1-\Delta^n_{\delta}) + 8\cdot 2^{-n \gamma }) - \tfrac{1}{n}.
\end{align}
Therefore such a random code has vanishing error probability for $n \to \infty$. As discussed earlier, at sufficiently large $n$ we have $\left\| \bar{\rho}'^{s}-\bar{\rho}^{s} \right\|_1 \leq \frac{1}{n}$, where $\Bar{\rho}^s = \mathds E_{X} \rho^s_{X}$. Thus with lemma \ref{thm:entropy_continuity} we have $S(\bar{\rho}'^{s}) \geq S(\bar{\rho}^{s}) - h\left(\tfrac{1}{n}\right) - E \cdot h\left(\tfrac{1}{E\cdot n}\right)$, implying
\begin{align}
    \lim_{n \to \infty} S(\bar{\rho}'^{s}) &= S(\bar{\rho}^{s}).
\end{align}

In the next part, we use the covering lemma (lemma \ref{thm:covering_lemma}) to show that the same random code achieves secrecy if we partition it into suitable ''fake`` ensembles. Thus we construct a set of projectors $\tilde{\Pi}^n$ and $\tilde{\Pi}^n_{x^n}$ with the properties
\begin{align}
\mathrm{Tr}\left[\tilde{\Pi}^n \tilde{\rho}'^s_{x^n}\right] &\geq 1-\tilde{\varepsilon},\label{eq:cover1}\\
\mathrm{Tr}\left[\tilde{\Pi}^n_{x^n} \tilde{\rho}'^s_{x^n}\right] &\geq 1-\tilde{\varepsilon},\label{eq:cover2}\\
\mathrm{Tr}\left[\tilde{\Pi}^n\right] &\leq \tilde{D},\label{eq:cover3}\\
\tilde{\Pi}^n_{x^n}\tilde{\rho}'^s_{x^n} \tilde{\Pi}^n_{x^n} &\leq \tfrac{1}{\tilde{d}} \tilde{\Pi}^n_{x^n},\label{eq:cover4}
\end{align}
for some $0<\Tilde{d}<\Tilde{D}$ and $0<\Tilde{\varepsilon}<1$. The code word projectors are simple, we use the fact that our states are pure states, therefore they are themselves rank-one projectors, and we can set $\tilde{\Pi}^n_{x^n} = \tilde{\rho}^s_{x^n}$, $\tilde{d} = 1$, fulfilling the properties \eqref{eq:cover2} and \eqref{eq:cover4}. For the code space projector, we use the strong typical subspace projectors corresponding to the average state $\tilde{\sigma}^s = \sum_{x \in \mathcal X}p_X(x) \tilde{\rho}'^s_x$ and set $\tilde{\Pi}^n = \Pi^n_{\tilde{\sigma}^s,\delta}$. This definition enables us to directly apply property 15.2.7 from \cite{wilde2017}, ensuring that \eqref{eq:cover1} is satisfied. To fulfill \eqref{eq:cover3} we set $\tilde{D} = 2^{n(S(\tilde{\sigma}^s)+\delta)}$, due to the properties of the typical subspace projector.

We have seen that for sufficiently large $n$ we have $\|\tilde{\sigma}^\eta - \tilde{\rho}^\eta \|_1 \leq \frac{1}{n}$, where $\tilde{\rho}^\eta = \mathds E_X \tilde{\rho}^\eta_X$ is the single-mode average state in infinite dimensions. With lemma \ref{thm:entropy_continuity} and the input energy constraint $E$ we see that the entropy of the approximated state approaches the entropy of the original state:
\begin{align} \label{eq:entropy_approx}
    S(\tilde{\sigma}^\eta) &\geq S(\tilde{\rho}^\eta) - h\left(\tfrac{1}{n}\right) - E\cdot h\left(\tfrac{1}{E\cdot n}\right).
\end{align}

Now let $\mathcal M$ be a set of messages and let $\mathcal L$ be an index set with $\abs{\mathcal M}\abs{\mathcal L}\leq K_n$. We slightly relabel our random code from before by selecting $\abs{\mathcal L}$ random variables $X_{m,l}^n$ from $X^n$ according to $p'_{X^n}(X_{m,l}^n)$ for each message $m \in \mathcal M$, giving us $\mathcal C_n = \left\{ X_{m,l}^n \right\}_{m \in \mathcal M, l \in \mathcal L}$. The average approximated state at the wiretapper of a specific code is 
\begin{align}
\tilde{\rho}'^s_\mathcal{C} = \frac{1}{\abs{\mathcal M}\abs{\mathcal L}}\sum_{m \in \mathcal M, l \in \mathcal L} \tilde{\rho}'^s_{X_{m,l}^n},
\end{align}
the average state of the ensemble defined by message $m$ is
\begin{align}
\tilde{\rho}'^s_{\mathcal{C},m} = \frac{1}{\abs{\mathcal L}}\sum_{l \in \mathcal L} \tilde{\rho}'^s_{X_{m,l}^n}. 
\end{align}
Then due to lemma \ref{thm:covering_lemma} we have
\begin{align}
\mathrm{Pr}\left\{ \|\tilde{\rho}'^s_\mathcal{C}-\tilde{\rho}'^s_{\mathcal{C},m}||_{1} \leq 30\tilde{\varepsilon}^{\frac{1}{4}} \right\} &\geq 1-2\tilde{D} 2^{-\tfrac{\varepsilon^{3}|\mathcal L|\tilde{d}}{4 \tilde{D}}},
\end{align}
given $\varepsilon$ is small and $|\mathcal L| \gg \frac{\tilde{\varepsilon}^{3} \tilde{d}}{\tilde{D}}$. Then for the non-approximated states
\begin{align}
\tilde{\rho}^s_\mathcal{C} =& \frac{1}{\abs{\mathcal M}\abs{\mathcal L}}\sum_{m \in \mathcal M, l \in \mathcal L} \tilde{\rho}^s_{X_{m,l}^n},\\
\tilde{\rho}^s_{\mathcal{C},m} =& \frac{1}{\abs{\mathcal L}}\sum_{l \in \mathcal L} \tilde{\rho}^s_{X_{m,l}^n},
\end{align}
we have with $\tilde{\varepsilon}=1/n^4$
\begin{align} \label{eq:privacy_criterion_proof}
\mathrm{Pr}\left\{ \|\tilde{\rho}^s_\mathcal{C}-\tilde{\rho}^s_{\mathcal{C},m}||_{1} \leq \tfrac{32}{n} \right\} &\geq 1-2\tilde{D} 2^{-\frac{n^{-12}|\mathcal L|\tilde{d}}{4 \tilde{D}}}.
\end{align}

Together with \eqref{eq:entropy_approx}, this ensures that the information leakage to the eavesdropper \eqref{eq:privacy} becomes arbitrarily small for sufficiently large $n$ if $\abs{\mathcal L} > 2^{nS(\tilde{\rho}^s)}$.
To ensure vanishing information leakage for all $s\in\mathbf S_n$ the legitimate parties therefore pick $\abs{\mathcal L} >  \sup_{s \in \mathbf S_n} 2^{nS(\tilde{\rho}^s)}$. Application of the union bound over all $s\in\mathbf S_n$ gives the desired result.

We have shown that our random code has high probability of achieving low error in data transmission and information leakage, with upper bounds approaching $0$ as $n$ goes to infinity. Due to the union bound, at least one code will achieve both, and we get for every $\varepsilon'$ and large enough $n$ the bound
\begin{align}
    \abs{\mathcal M} > \inf_{s \in \mathbf S_n} \, 2^{n(S(\Bar{\rho})-S(\tilde{\rho}^s) -\varepsilon')}.
\end{align}
Since $\varepsilon'$ is arbitrary this implies
\begin{align}
    C_\mathrm{noCSI} \geq& \lim_{n\to\infty}\inf_{s \in \mathbf S_n} S(\Bar{\rho}^s)- \sup_{s \in \mathbf S_n}S(\tilde{\rho}^s).
\end{align}
In the case of channel state information at the sender, the transmission is cut in two blocks of length $n_1=\sqrt{n}$ and $n_2=n-n_1$. In the first block, the sender prepends an attenuator channel such that the actual channel has transmissivity $\tau_{\min}:=(\tau_{\min},\eta') \in \mathbf S_n : \tau_{\min} \leq \tau \forall (\tau,\eta) \in \mathbf S_n $. The sender and receiver then use a data transmission code for the pure loss channel of transmissivity $\tau_{\min}^2$ with capacity $C''>0$, to transmit $M_1:=2^{\sqrt{n}C''}$ bits of the channel state information, thereby describing two sets $(\tau_a,\tau_b]$ and $(\eta_a,\eta_b]$ with the properties $\tau_a\leq\tau\leq\tau_b$ and $\eta_a\leq\eta\leq\eta_b$ and $|\tau_a-\tau_b|\leq 2^{-M_1}$ and $|\eta_a-\eta_b|\leq2^{-M_1}$. Afterwards they use a code with no CSI for the compound channel with state set $\mathbf S_n':=(\tau_a,\tau_b]\times(\eta_a,\eta_b]$, thereby establishing 
\begin{align}
    C_\mathrm{CSI} \geq& \lim_{n\to\infty}\inf_{s \in \mathbf S_n'} \left( S(\Bar{\rho}^s)- S(\tilde{\rho}^s)  \right)-\varepsilon'
\end{align}
where $\varepsilon'$ can be made as small as possible. The probability distribution $p(u)$ that maximizes the entropy of the state $\int_{u \in \mathbb C} p(u) \mathcal N_\tau(u) du$ under the energy constraint $\Tr[\hat{N}\int_{u \in \mathbb C} p(u) \mathcal N_\tau(u) du] \leq E$ is the complex Gaussian distribution $p(u)=\frac{1}{\pi E}2^{-\left| u \right|^{2}/E}$, with entropy
\begin{align}
    S\big( \int_{u \in  \mathbb C} p(u) \mathcal N_\tau(u) du \big) = g(\tau^2 E).\label{eq:entropy-of-gaussian-ensemble}
\end{align}
We use discretizations of this distribution to prove Theorem \ref{thm:main} as follows:
We consider ensembles $\mathcal E_\Delta = \left\{ \ketbra{x}, p^\Delta_X(x) \right\}_{x \in \mathcal X_\Delta}$, where for $\Delta > 0$, $\mathcal X_\Delta$ is a finite set, which approximate the complex Gaussian in the sense that
\begin{align}
\max_{0\leq\gamma\leq1}\left\|\frac{1}{\pi E} \int_{\mathbb{C}}2^{-\left| x \right| ^{2}/E}\mathcal N_\gamma(x)dx-{\nu}^\gamma_\Delta \right\|_1 &\leq \Delta
\end{align}
where ${\nu}_\Delta^\gamma = \sum_{x \in \mathcal X_\Delta} p^\Delta_X(x) \mathcal{N}_\gamma(x)$ and $\Tr[\hat{N}\nu^\gamma_\Delta] \leq E$ for all $\gamma$. Details are in the Appendix. Then lemma \ref{thm:entropy_continuity} implies
\begin{align}
\lim_{\Delta \to 0} S(\nu^1_{\Delta}) = S\left( \int_{u \in \mathbb C} p(u) \mathcal N_1(u) du \right).
\end{align}
Due to \eqref{eq:entropy-of-gaussian-ensemble}, we have thus proven that
\begin{align}
    C_\mathrm{CSI} =& \lim_{n\to\infty}\inf_{s \in \mathbf S_n} \big( g(\tau^2E)-g(\eta^2E) \big)\\
    C_\mathrm{noCSI} =& \lim_{n\to\infty}\big(\inf_{s \in \mathbf S_n} g(\tau^2E)- \sup_{s \in \mathbf S_n}g(\eta^2E)\big)_+.
\end{align}
Lastly we specify $\mathbf S_n$. For any $\mathbf{S} \subset [0,1]\times[0,1]$ there is a $\mathbf{S}'_{\mu^{-1/2}}\subset \mathbf S$ such that $\forall (\tau,\eta)\in \mathbf{S}\ \exists(\tau',\eta')\in \mathbf{S}'_{\mu^{-1/2}}:\left| \tau'-\tau \right|\leq \mu,\left| \eta'-\eta \right|\leq \mu$ and $\left| \mathbf{S}'_{\mu^{-1/2}} \right|\leq \frac{1}{\mu^{2}}$. We can simplify the argument by realizing that in the discretization $\mathcal{X}_\Delta$ there is an energy cutoff $\hat{E}_\Delta = \max_{x \in \mathcal{X}_\Delta}\left| x \right|^{2}$. Further, the trace distance of two coherent states can be given explicitly as $\||\alpha\rangle\langle\alpha|-|\beta\rangle\langle\beta|\|_1 = 2\cdot\sqrt{1-e^{-\abs{\alpha-\beta}^2}}$, which implies that for every $x^n \in \mathcal{X}_\Delta^n$ and $s \in \mathbf{S}$ there is $s' \in \mathbf{S}'_{\mu^{-1/2}}$ such that $\left\| \rho_{x^n}^{s'}- \rho_{x^n}^s \right\|_1\leq 2\sqrt{ 1-e^{-n\mu\hat{E}_\Delta} }$. With $\mu = \frac{1}{n^{2}}$ and lemma \ref{thm:fin_sup} we find for every $\mathbbm1\geq X\geq0$
\begin{align}
\mathrm{Tr}\left[\mathcal{N}_{s'}^{\otimes n}(x^n)X\right]\geq \mathrm{Tr}\left[\mathcal{N}_{s}^{\otimes n}(x^n)X\right] - 4\sqrt{\hat{E}/n}.
\end{align}
Since $|\mathbf S'_{n}|=\mathcal O(n^4)$ and $\mathbf S'_n\subset\mathbf S$, Theorem \ref{thm:main} is proven. 

\section*{Appendix}
\subsection*{Gaussian Discretization}
We wish to discretize a Gaussian ensemble of coherent states $\left\{ \frac{1}{\pi E}e^{-\left| x \right|^{2}/E}, \ketbra{x}{x} \right\}_{x \in \mathbb{C}}$ with average state $\bar{\rho}=\int_{x \in \mathbb{C}} \frac{1}{\pi E}e^{-\left| x \right|^{2}/E} \ketbra{x}{x}dx$ in such a way that we get an ensemble $\left\{p_{X}(x), \ketbra{x}{x} \right\}_{x \in \mathcal{X}}$, where $\mathcal{X}$ is a finite set and $p_{X}$ a probability distribution with average state $\bar{\rho}' = \sum_{x \in \mathcal{X}}p_{X}(x)\ketbra{x}{x}$ close to the original and bounded energy:
\begin{align}
\| \bar{\rho}'-\bar{\rho} \|\leq \varepsilon, \qquad \mathrm{Tr}[\hat{N}\bar{\rho}']\leq \mathrm{Tr}[\hat{N}\bar{\rho}], 
\end{align}
for some $\varepsilon>0$, where $\hat{N}$ is the photon number operator.

We partition the complex plane into a finite number of patches and assign each patch a complex number $x$ and a probability $p_{X}(x)$. The probability assigned to a patch $T_{x} \subset \mathbb{C}$ is $p_{X}(x) = \int_{z \in T_{x}}\frac{1}{\pi E}e^{-\left| z \right|^{2}/E}dz$ and we pick the corresponding $x$ such that $x \in P_{x}$ and$\left| x \right|^{2} = \mathrm{Tr}\left[\hat{N}\left( \int_{z \in T_{x}}\frac{1}{\pi E}e^{-\left| z \right|^{2}/E}\ketbra{z}{z}dz \right)\right]$. This does not fully define $x$, but it ensures that the average energy of the discretized ensemble does not exceed the average energy of the continuous ensemble and is sufficient for our purposes. We first consider everything within a circle of radius $R$ around the origin. For any $0<r\leq R$ this part of the complex plane can be partitioned in such a way that each patch is contained in a disk of radius $r$, by using $c\left( \frac{R}{r} \right)^{2}$ patches, for some constant $c$. This implies that for any $x' \in P_{x}, \left| x-x' \right|\leq 2r$ and therefore $\left\| \ketbra{x'}{x'} - \ketbra{x}{x} \right\|_{1} \leq 2\sqrt{ 1-e^{-4r^{2}} }$. The same bound applies to the average state of any patch:
\begin{align}
&\left\| \ketbra{x}{x}- \int_{z \in T_{x}}\frac{1}{\pi E}e^{-\left| z \right|^{2}/E}\ketbra{z}{z}dz\right\|_{1} \\
\leq&  \int_{z \in T_{x}} \frac{1}{\pi E}e^{-\left| z \right|^{2}/E} \left\| \ketbra{z}{z} - \ketbra{x}{x} \right\|_{1}dz \\
\leq & \int_{z \in T_{x}} \frac{1}{\pi E}e^{-\left| z \right|^{2}/E}  2\sqrt{ 1-e^{-4r^{2}} } dz\\
=  & 2 p_{X}(x) \sqrt{ 1-e^{-4r^{2}} }.
\end{align}
The the remaining complex plane outside the disk of radius $R$, we simply assign value zero: $p_{X}(0)=\int_{\left| x \right|>R}\frac{1}{\pi E}e^{-\left| x \right|^{2}/E}dx=e^{-R^{2}/E}$, or if there was already a positive probability assigned to zero, we add them. Then
\begin{align}
\left\| \bar{\rho}-\bar{\rho}' \right\| _{1} \leq 2 (1-e^{-R^{2}/E}) \sqrt{ 1-e^{-4r^{2}} } + 2e^{-R^{2}/E}
\end{align}
By picking suitable $r$ and $R$, we achieve the desired result.

\subsection*{Coding Lemma Modification}
Lemma \ref{thm:compound_rate} is a modified version of lemma 1 in \cite{Bjelakovic2013}. The proof makes use of the states $\rho_n$ and $\tau_n$. In order to prove this version we define slightly different states
\begin{align}
\rho'_{n} =& \frac{1}{\left| \mathbf{S}_n \right| }\sum_{s \in \mathbf{S}_n} \sum_{x^n \in \mathbf{X}^n} p_{n}'(x^n) \delta_{x^n} \otimes \mathcal N_{s}^{\otimes n}(x^n),\\
\tau'_{n} =& \sum_{x^n \in \mathbf{X}^n} p'_{n}(x^n) \delta_{x^n} \otimes\frac{1}{\left| \mathbf{S}_n \right| }\sum_{s \in \mathbf{S}_n}\sum_{x^n \in \mathbf{X}^n} p'_{n}(x^n)\mathcal N_{s}^{\otimes n}(x^n)\nonumber
\end{align}
where $\delta_{x^n}:=\ket{e_{x^n}}\bra{e_{x^n}}$, $\ket{e_{x^n}}:=\otimes_{i=1}^n\ket{e_{x_i}}$ and $\ket{e_{x_i}}$ form an orthonormal basis of $\mathbb C^{|\mathbf X|}$. We then follow the steps in \cite{Bjelakovic2013} with minor modifications. We observe that $\|\rho_{n}-\rho_{n}' \|_{1}=2(1-\Delta_{\delta}^n)$.
Then with lemma \ref{thm:fin_sup} we have 
\begin{align}
\mathrm{Tr}\left[P_{n} \rho'_{n}\right] \geq 1-\lambda-2(1-\Delta_{\delta}^n),
\end{align}
For the second property, $\mathrm{Tr}\left[P_{n} \tau_{n}\right] \leq 2^{-na}$, consider that if we replace $p'_{n}(x^n)$ in the definition of $\tau_{n}'$ with $\Delta_{\delta}^{n}p'_{n}(x^n)$, then the only difference to $\tau_{n}$ is the removal of positive terms, and therefore $\tau'_{n}\leq \frac{1}{(\Delta_{\delta}^{n})^{2}}\tau_{n}$. This implies that
\begin{align}
\mathrm{Tr}\left[P_{n}\tau'_{n}\right]\leq \frac{1}{(\Delta_{\delta}^{n})^{2}} \mathrm{Tr}\left[P_{n}\tau_{n}\right]\leq \frac{1}{(\Delta_{\delta}^{n})^{2}}2^{-na}.
\end{align}
The rest of the proof is equivalent to that in \cite{Bjelakovic2013}.
\bibliographystyle{plain}
\bibliography{bib}
\end{document}